\documentclass[a4paper,11pt]{article}
\pdfoutput=1 

\usepackage{jcappub} 
\usepackage[export]{adjustbox}
\usepackage{subfigure} 

\usepackage[T1]{fontenc} 

\newcommand{\beq}{\begin{equation}}
\newcommand{\eeq}{\end{equation}}
\newcommand{\bea}{\begin{eqnarray}}
\newcommand{\eea}{\end{eqnarray}}
\newcommand{\bi}{\begin{itemize}}
\newcommand{\ei}{\end{itemize}}
\newcommand{\bfi}{\begin{figure}[!t]
\epsfxsize=9cm
\epsffile}
\newcommand{\bfig}{\begin{figure*}[!t]
\center{}
\epsfxsize=15cm
\epsffile}
\newcommand{\efi}{\end{figure}}
\newcommand{\efig}{\end{figure*}}
\newcommand{\no}{\nonumber}
\newcommand{\mpch}{{\rm Mpc}/h}
\newcommand{\hmpc}{h/{\rm Mpc}}

\newcommand{\bfk}{{\bf k}}
\newcommand{\bfv}{{\bf v}}
\newcommand{\btau}{\bar{\tau}}
\newcommand{\bfs}{{\bf s}}

\newcommand{\thetac}{\theta_{\rm c}}

\newcommand{\msun}{{\rm M}_\odot}
\newcommand{\muKa}{\mu K\mathchar`-{\rm arcmin}}
\newcommand{\ZY}[1]{{\color{black} #1}}

\title{\boldmath Forecast constraints on the baryonic feedback effect from the future kinetic Sunyaev-Zel'dovich effect detection}

\author[a,b]{Yi Zheng}
\author[c,d,e]{and Pengjie Zhang}

\affiliation[a]{School of Physics and Astronomy, Sun Yat-sen University, 2 Daxue Road, Tangjia, Zhuhai, 519082, China}
\affiliation[b]{CSST Science Center for the Guangdong-Hong kong-Macau Greater Bay Area, SYSU, Zhuha, 519082, China}
\affiliation[c]{Department of Astronomy, School of Physics and Astronomy, Shanghai JiaoTong University, Shanghai 200240, China}
\affiliation[d]{Tsung-Dao Lee Institute, Shanghai Jiao Tong University, Shanghai, 200240, P.R.China}
\affiliation[e]{Key Laboratory for Particle Astrophysics and Cosmology (MOE)/Shanghai Key Laboratory for Particle Physics and Cosmology, China}

\emailAdd{zhengyi27@mail.sysu.edu.cn}

\abstract{The baryonic feedback effect is an important systematic error in the weak lensing (WL) analysis. It contributes partly to the $S_8$ tension in the literature. With the next generations of large scale structure (LSS) and CMB experiments,  the high signal-to-noise kinetic Sunyaev-Zel'dovich (kSZ) effect detection can tightly constrain the baryon distribution in and around dark matter halos, and quantify the baryonic effect in the weak lensing statistics. In this work, we apply the Fisher matrix technique to predict the future kSZ constraints on 3 kSZ-sensitive Baryon Correction Model (BCM) parameters. Our calculations show that, in combination with next generation LSS surveys, the 3rd generation CMB experiments such as AdvACT and Simon Observatory can constrain the matter power spectrum damping $S(k)$ to the precision of $\sigma_S(k)<0.8\%\sqrt{37.8{\rm Gpc}^3h^{-3}/V}$ at $k\lesssim 10\hmpc$, where $V$ is the overlapped survey volume between the future LSS and CMB surveys. For the 4th generation CMB surveys such as CMB-S4 and CMB-HD, the constraint will be enhanced to $\sigma_S(k)<0.4\%\sqrt{37.8{\rm Gpc}^3h^{-3}/V}$. If extra-observations, e.g. X-ray detection and thermal SZ observation, can effectively fix the gas density profile slope parameter $\delta$, the constraint on $S(k)$ will be further boosted to $\sigma_S(k)<0.3\%\sqrt{37.8{\rm Gpc}^3h^{-3}/V}$ and $\sigma_S(k)<0.1\%\sqrt{37.8{\rm Gpc}^3h^{-3}/V}$ for the 3rd and 4th generation CMB surveys.}
\begin{document}
\maketitle
\flushbottom


\section{Introduction}
\label{sec:Intro}

Weak gravitational lensing (WL) serves as the most important cosmic probe in the current era of the large scale structure cosmology~\citep{Bartelmann2001,Hoekstra2008,Kilbinger2015}. By measuring the projected matter density fields weighted by cosmic distances, the WL effect is simultaneously sensitive to the cosmic expansion and structure growth histories (e.g.~\cite{Albrecht2006}). In particular, it directly maps the density fluctuation of all matters, avoiding the complicated bias issues associated with probes like the galaxy clustering (GC)~\cite{Desjacques18biasreview}. The high precision of the WL detection expected from the stage IV galaxy imaging surveys
drives the demands for the WL theory accuracy to an unprecedented level (e.g.~\cite{Amendola2018,GongY2019}). This paper is dedicated to contribute to this issue.

By the Limber approximation~\citep{Limber1953}, angular statistics from the WL effect in Fourier space are expressed as integrals of the weighted 3-dimensional matter density power spectrum along the line-of-sight (LOS). At non-linear scales, the matter power spectrum is damped due to the baryonic feedback effect, in that the baryon evolution rearranges the matter distribution in and around dark matter halos (e.g.~\cite{Semboloni2011,Harnois-Deraps2015}). 
Since the damping can reach beyond $10\%$ at $k>1\hmpc$, this baryonic effect is regarded as one of the most severe systematic errors in the WL analysis. Studies have shown that it contributes partly to the $S_8$ tension~\cite{Schneider2022,ChenA2023,Arico2023,McCarthy2023}. It is challenging to quantify this damping from the first principle evaluation, e.g., by the hydrodynamical simulation, as different phenomenological sub-grid physics implied by various hydro-simulation suits induce a large difference of the damping levels (e.g.~\cite{Schneider2022}). Moreover, the heavy numerical resources required by running  hydro-simulations makes impossible the direct inclusion of the sub-grid physics into the Bayesian analysis of the WL data. Therefore, the small scale WL data are sometimes discarded during the cosmological WL analysis, and careful tests are carried out to make sure that the scale cuts are conservative enough to avoid significant baryon-effect-induced systematics (e.g.~\cite{Amon2022}). 

Efforts have been made to alleviate this obstacle. First, alternative methods have been developed to approximately describe the baryonic effects on the non-linear matter power spectrum. Typical examples include (1) \texttt{HMcode}, a modified halo model taking into account the baryonic effects on the matter power spectrum level~\cite{Mead2015,Mead2021}; and (2) the `Baryonification' method, which mimics the baryoinc effects on the simulation field level by adjusting the radial positions of dark matter particles within halos according to the `Baryon Correction Model (BCM)'~\cite{Schneider2015}. Instead of the direct simulation of baryons, these methods can mimic the barynoic effects in a much quicker and cheaper way. 

Second, emulators concerning the baryonic effects have been developed. While \texttt{HMcode} can be directly incorporated into a MCMC analysis and make inference of cosmological and astrophysical parameters, the implementation of BCM needs an emulator to effectively span a wide parameter space. Two mature emulators have be developed by~\cite{Giri2021} and~\cite{Arico2021}, in which the \texttt{Baccoemu} emulator~\cite{Arico2021} has been applied on the small scale DES data to simultaneously constrain cosmology and the baryon feedback~\citep{ChenA2023,Arico2023}. In addition, the \texttt{SP(k)} emulator directly based on a large hydro-simulation suit has also been developed~\cite{Salcido2023}, and it emphasizes the tight relation between the group baryon fraction and  the non-linear matter power spectrum damping.

Third and closely related to the theme of this work, additional observations are incorporated to constrain the small scale baryon effects, by which we  efficiently break the degeneracy between cosmological and astrophysical parameters. The extra information can come from observations of the X-ray, thermal Sunyaev-Zel'dovich (tSZ) and kinetic Sunyaev-Zel'dovich (kSZ) effects et al., which separately targets the baryonic temperature, pressure and density distributions on halo scales~\cite{Schneider2022,Grandis2023}. With combinations of all these measurements including the WL effect, and with accurate and compact models or emulators of the baryonic effect, it is promising to pin down the baryonic backreaction and avoid the associated systematic cosmological parameter biases in a WL analysis~\cite{Schneider2020b}.

This prospect can be realized by the next generation of LSS and CMB surveys. The next generation of LSS surveys, such as DESI~\cite{DESI2016}, Euclid~\cite{Euclid}, LSST~\cite{LSST2012} and CSST~\cite{GongY2019}, will provide one order of magnitude larger size of the tracer number and the survey volume, likely producing a large volume-limited galaxy cluster catalog with the cluster mass $M>10^{13}M_\odot/h$. The next generation of CMB surveys, such as AdvACT~\cite{Farren2023}, the Simon Observatory (SO)~\cite{Simon2019}, CMB-S4~\cite{CMBS42019} and CMB-HD~\cite{CMBHD2020}, will provide wide CMB maps with several times to one order of magnitude smaller \ZY{detector} noise and with an angular resolution down to $1'$ or lower.  All these improvements promise to consolidate our understanding of the universal baryon distribution and make credible the small scale WL analysis. To strengthen our confidence of this roadmap, in this paper we will take the kSZ measurement as an example and portrait this improvement in a quantitative level. Adopting the fisher matrix technique, we will show how the one-order-of-magnitude improvement of the kSZ detection S/N in the near future~\cite{Sugiyama18} will narrow the matter power spectrum damping uncertainty  down to $<1\%$ at $k>1\mpch$, if the degeneracy between BCM parameters are appropriately broken.

This paper will be organized as follows. In section~\ref{sec:theory} we introduce the BCM and the way we calculate the observed kSZ signal. In section~\ref{sec:mock_obs} we construct the mock kSZ observations that will be utilized in section~\ref{sec:results} to constrain the baryonic effect described by the BCM. Conclusions are made in section~\ref{sec:conclusion}. Throughout this paper, we will assume a WMAP 9 cosmology~\citep{WMAP9}, such as $\{\Omega_{\rm b}, \Omega_{\rm CDM}, h, n_s, \sigma_8 \} = \{0.046, 0.235, 0.697, 0.971, 0.82\}$. We do not expect that the main conclusion heavily depend on the choice of the cosmology.


\section{Theory}
\label{sec:theory}


The kSZ effect is a secondary CMB anisotropy in which the temperature of CMB photons is slightly changed by their scattering off of free electrons with a bulk motion~\citep{kSZ1970,kSZ1972,kSZ1980}. It detects the momentum field of free electrons in our Universe, as illustrated by 
\beq
\label{eq:deltaT_kSZ}
\frac{\delta T_{\rm kSZ}(\hat{n})}{T_0} = -\sigma_{\rm T}\int dl n_{\rm e} \left(\frac{{\bfv}_{\rm e}\cdot \hat{n}}{c}\right)  \,.
\eeq
Here $T_0\simeq 2.7255\rm K$ is the averaged CMB temperature, $\sigma_{\rm T}$ is the Thomson-scattering cross-section, $c$ is the speed of light, $\hat{n}$ is the unit vector along the line of sight (LOS), $n_{\rm e}$ is the physical free electron number density, $\bfv_{\rm e}$ is the proper peculiar velocity of free electrons, defined to be positive for those recessional objects, and the integration $\int dl$ is along the LOS given by $\hat{n}$.

With the assumption that most CMB photons encounter only one large halo during their journey to us, eq.~(\ref{eq:deltaT_kSZ}) can be reformed as 
\beq
\label{eq:deltaT_kSZ2}
\frac{\delta T_{\rm kSZ}(\hat{n}_i)}{T_0} = -\frac{\tau_{i}}{c}\bfv_i\cdot\hat{n}_i\approx-\frac{\btau}{c}\bfv_i\cdot\hat{n}_i\,,
\eeq
where $\tau_{i}=\int dl\sigma_{\rm T}n_{{\rm e},i}$ is the optical depth of the $i$th halo and $\btau$ is the average optical depth of a halo sample.

In this work we will apply the Fisher matrix technique to the mock kSZ observation and study its constraining power on the baryonic feedback effect. We select halos from N-body simulations and theoretically portray the gas/free-electron distribution in and around halos. We obtain the optical depth $\bar{\tau}$ by integrating the free electron number density along the LOS and then we convolve $\bar{\tau}$ with the CMB beam function and the aperture photometer (AP) filter to calculate the observed $\bar{\tau}_{\rm obs}$. The gas distribution within halos will be described via the ‘Baryon Correction Model (BCM)', introduced in section~\ref{subsec:gas_profile}, and the convolution will be introduced in section~\ref{subsec:tau_theory}.

\subsection{Gas density profile from the Baryon Correction Model}
\label{subsec:gas_profile}
\begin{table}[t!]
\centering
\begin{tabular}{ccccc}
\hline\hline
$\Theta$ & $\Theta_{\rm mid}$ & $\Theta_{\rm S22}$ & $\Theta_{\rm G23}$ & Range  \\
\hline
${\log_{10} M_{\rm c}}^\ast$ & {13} & {13.2} & {14.53} &{[11,15]} \\
${\theta_{\rm ej}}^\ast$ & {5} & {5} & {4.36} & {[2,8]} \\
${\delta}^\ast$ & {7} & {7} & {6} & {[3,11]} \\
$\gamma$ & 2.5 & 2.5 & 1.92 & [1,4]\\
$\eta_{\rm star}$ & 0.225 & 0.24 & 0.23 & [0.05,0.4] \\
$\mu$ & 1 & 0.3 & 0.5 & [0,2] \\
\hline\hline
\end{tabular}
\caption{Fiducial BCM parameter values and their allowed varying ranges in the \texttt{BCemu} emulator. Parameters with a $\ast$ superscript are kSZ-sensitive ones that we analyze with the Fisher matrix technique in this paper, and their varying ranges are also regarded as flat priors in the Fisher matrix analysis. In the 6 parameters, $\log_{10} M_{\rm c}$ and $\mu$ determine the gas profile slope parameter $\beta$ by eq.~(\ref{eq:beta}). $\delta$ and $\gamma$ also affect the slope of the gas density profile. $\theta_{\rm ej}$ controls the characteristic radius of the ejected gas distribution, and $\eta_{\rm star}$ determines the star fraction of a halo vis eq.~(\ref{eq:f_star}). The default parameter fiducial values $\Theta_{\rm mid}$ are chosen to be at centers of the varying ranges, and we also try the best fitted BCM parameters in S22~\cite{Schneider2022} and G23~\cite{Grandis2023} as fiducial values.}
\label{table:para}
\end{table}

According to the BCM, instead of running a hydro-simulation, the Baryonification methodology mimics the baryonic feedback on the matter distribution in a dark matter only simulation by adjusting the radial positions of dark matter particles within halos. Phenomenological parameters in the BCM describe the final halo density profile after the dark matter-baryon co-evolution, which is generally decomposed into three components~\cite{Giri2021}:
\beq
\label{eq:rho_decompose}
\rho(r) = \rho_{\rm DM} + \rho_{\rm gas} + \rho_{\rm star}\,,
\eeq
namely a dark matter profile $\rho_{\rm DM}$, a gas profile $\rho_{\rm gas}$ and a stellar component profile $\rho_{\rm star}$. The kSZ observation is directly sensitive ot $\rho_{\rm gas}$, and its modelling is detailed as follows.

The version of the BCM we choose is based on~\cite{Schneider2019,Giri2021} (\texttt{S19} model hereafter). It is modified from the original BCM paper~\cite{Schneider2015}, and is different from the \texttt{Bacco}-model developed in~\cite{Arico2021}. The two models are similar to each other, in that they have the same number of free parameters, and have similar constraining power when confronted to the real observations~\cite{Grandis2023}. So we do not expect that the choice of the BCM version will bias the main conclusion of this work. 

In the \texttt{S19} model, the gas profile is parameterised as
\beq
\label{eq:gas_profile}
\rho_{\rm gas}(r) = \frac{\rho_{\rm gas,0}}{(1+u)^\beta(1+v^\gamma)^{(\delta-\beta)/\gamma}}\,,
\eeq
where $\gamma$ and $\delta$ are set to be two free parameters, $u\equiv r/r_{\rm co}$ and $v\equiv r/r_{\rm ej}$. 
The two characteristic radii are defined as
\beq
\label{eq:gas_radii}
r_{\rm co}\equiv \theta_{\rm co}r_{\rm 200c}\,,\,\,\,\,\,\,r_{\rm ej}\equiv \theta_{\rm ej}r_{\rm 200c}\,,
\eeq
in which $\theta_{\rm co}=0.1$ and $\theta_{\rm ej}$ is a free model parameter controlling the characteristic radius of the ejected gas distribution. {As we can see, $\delta$ indicates the density slope of the ejected gas at outer halo regions, while $\gamma$ controls the transition between the inner region density slope to the outer region slope, which is a second order effect in quantifying the gas density profile. \ZY{This is verified in figure~\ref{fig:tau_sensitivity} of appendix~\ref{app:tau_BCM_dependency}, where we notice that $\bar{\tau}_{\rm obs}$ is more sensitive to $\beta$ than $\gamma$ at inner halo regions, and is more sensitive to $\delta$ than $\gamma$ at outer halo regions.}

Another slope parameter of the gas profile, $\beta$, is parameterised as
\beq
\label{eq:beta}
\beta(M_{\rm 200c}) = \frac{3(M_{\rm c}/M_{\rm 200c})^\mu}{1+(M_{\rm c}/M_{\rm 200c})^\mu}\,, 
\eeq
where $M_{\rm c}$ and $\mu$ are two free parameters. $\beta$ determines the inner halo region gas density. It is always positive by definition, and approaches 3 when $M_{\rm c}\rightarrow\infty$. As shown in eq.~(\ref{eq:beta}), \ZY{$M_{\rm c}$ is the main parameter that determines the $\beta$ value, }
while $\mu$ is sub-dominant and characterizes the speed of $\beta$ variation in terms of $M_{\rm c}$. \ZY{This is also verified in figure~\ref{fig:tau_sensitivity}, where $\bar{\tau}_{\rm obs}$ is shown to be more sensitive to $M_{\rm c}$ than $\mu$. }

The normalization parameter $\rho_{\rm gas,0}$ is given by 
\beq
\label{eq:rho_gas0}
\rho_{\rm gas,0} = f_{\rm gas}M_{\rm tot}\left[4\pi\int_0^\infty dr\frac{r^2}{(1+u)^\beta(1+v^\gamma)^{(\delta-\beta)/\gamma}}\right]^{-1}\,,
\eeq
\ZY{where $f_{\rm gas} = \Omega_b/\Omega_m-f_{\rm star}$ is the gas fraction of the whole halo and $M_{\rm tot}$ is the total halo mass\footnote{\ZY{When $r\rightarrow\infty$, the denominator of the integrand in eq.~(\ref{eq:rho_gas0}) tends to be $\propto r^\delta$. As such, the integration in eq.~(\ref{eq:rho_gas0}) is not always convergent for all $\delta$ values. For example, on the top right panel of figure~\ref{fig:tau_sensitivity}, we omit the case when $\delta=3$ due to the divergence of this integration.}}. 

Observationally, the observed gas fraction $f^{\rm obs}_{\rm gas}(r)\equiv \bar{\rho}_{\rm gas}(r)/\bar{\rho}_m(r)$ is always restricted within a certain halo radius $r$. Here $\bar{\rho}_{\rm gas}(r)$ and $\bar{\rho}_m(r)$ are the averaged densities of the gas and matter within a certain $r$. Since the halo has different matter, gas, and star density profiles,  $f^{\rm obs}_{\rm gas}(r)$ should be a function of radius. As we can see, the \texttt{S19} model enforces the universal baryon fraction $f_b\equiv \Omega_b/\Omega_m$ to all halos. Therefore, the $f_{\rm gas}$ here is $f^{\rm obs}_{\rm gas}(r)$ when $r\rightarrow\infty$. We will need to recalculate $f^{\rm obs}_{\rm gas}(r)$ when we are about to compare the theoretical result with the observed one.}

The stellar fraction $f_{\rm star}$ can be given by
\beq
\label{eq:f_star}
f_{\rm star}(M_{\rm 200c}) = A\left(\frac{M_1}{M_{\rm 200c}}\right)^{\eta_{\rm star}}\,,
\eeq
with $A=0.09$ and $M_1=2.5\times 10^{11}M_\odot/h$. $\eta_{\rm star}$ is a free parameter which can be robustly constrained by the optical data~\citep{Grandis2023}. The total halo mass is calculated by
\beq
\label{eq:M_tot}
M_{\rm tot} = 4\pi\int_0^{\infty} dr r^2 \rho_{\rm nfw}(r)\,,
\eeq
with $\rho_{\rm nfw}(r)$ being a truncated NFW profile
\beq
\label{eq:nfw_trunc}
\rho_{\rm nfw}(x) = \frac{\rho_{\rm nfw,0}}{x(1+x)^2}\frac{1}{(1+y^2)^2}\,,
\eeq
where $x\equiv r/r_s$ and $y\equiv r/r_t$. The scale radius $r_s$ is connected to $r_{\rm 200c}$ via the halo concentration $c\equiv r_{\rm 200c}/r_s$, and $r_t\equiv \epsilon\times r_{\rm 200c}$ with $\epsilon = 4$. For the NFW profile, the halo concentration is related to the halo mass $M_{\rm 200c}$ via~\cite{Cooray02}
\beq
\label{eq:c_M}
c(M_{\rm 200c},z) = \frac{9}{1+z}\left[\frac{M_{\rm 200c}}{M_\ast(z)}\right]^{-0.13}\,, 
\eeq
where $M_\ast(z)$ is the characteristic mass scale at which $\nu(M,z)$ = 1.

\ZY{In summary, there are 6 free parameters describing $\rho_{\rm gas}$ of the \texttt{S19} model. Within them, $\eta_{\rm star}$ can be determined by the optical data; $\beta$ determines the inner region distribution of $\rho_{\rm gas}$, which itself depends mainly on $M_{\rm c}$; $\delta$ determines the $\rho_{\rm gas}$ profile at large radii, and $\gamma$ characterizes the subtle transition of the density profile slope from inner to outer regions of the halo. In turn, we consider  $\gamma$ and $\mu$ to be sub-dominant in characterizing $\rho_{\rm gas}$, and  there remain 3 parameters that are most sensitive to the kSZ effect. We will set free these 3 parameters in our Fisher matrix analysis and fix other ones to their fiducial values. 

As mentioned above, this choice of free parameters choice is supported by the sensitivity test of $\bar{\tau}_{\rm obs}$ in appendix~\ref{app:tau_BCM_dependency}. Yet, in figure~\ref{fig:tau_sensitivity}, considerable degeneracy of $\gamma$ and $\mu$ with other parameters is also observed. By fixing these two parameters, we are effectively considering a joint analysis between the kSZ observation and extra datasets, such as optical data, tSZ and X-ray observations to break the degeneracy and pin down the uncertainties of $\eta_{\rm star}$, $\gamma$ and $\mu$.

The default fiducial values of 6 parameters and their varying ranges in the \texttt{BCemu} emulator are listed in table~\ref{table:para}. We also set the fiducial values to the best fitted BCM parameters in S22~\cite{Schneider2022} and G23~\cite{Grandis2023} to test the sensitivity of our conclusions to the choice of the fiducial values.}

\subsection{Optical depth detected by the kSZ effect}
\label{subsec:tau_theory}

The projected gas density profile
\beq
\label{eq:rho_gas_prj}
\rho^{\rm prj}_{\rm gas}(r) = 2\int^\infty_0\rho_{\rm gas}(\sqrt{r^2+l^2})dl\,
\eeq
is related to the theoretical optical depth by
\bea
\label{eq:tau}
\bar{\tau}(\theta) &=& 2\sigma_{\rm T}\int^\infty_0 n_{\rm e}(\sqrt{d_A(z)^2\theta^2+l^2})dl \\
&=& \frac{X_H+1}{2m_{\rm amu}}\sigma_T\rho_{\rm gas}^{\rm prj}(d_A(z)\theta)\,,
\eea
in which  $d_A(z)$ is the angular diameter distance, and the electron number density profile $n_{\rm e}(r)$ is related to $\rho_{\rm gas}(r)$ via
\beq
\label{eq:n_e}
n_{\rm e}(r) = \frac{X_H+1}{2m_{\rm amu}}\rho_{\rm gas}(r)\,,
\eeq
with the hydrogen mass fraction $X_H=0.76$ and the atomic mass unit $m_{\rm amu}=1.66\times10^{-27}$kg.

The observed optical depth $\bar{\tau}_{\rm obs}$ is evaluated by the convolution between $\tau(\theta)$, the CMB beam function, and the AP filter by~\footnote{The 2-D Fourier transform is defined as $f(\vec{\theta})=\int\frac{d^2l}{(2\pi)^2}e^{i\vec{l}\cdot\vec{\theta}}f(\vec{l})$, $f(\vec{l})=\int d^2\theta e^{-i\vec{l}\cdot\vec{\theta}}f(\vec{\theta})=2\pi\int_0^\infty f(\theta)J_0(l\theta)\theta d\theta$.}~\cite{Sugiyama18}
\beq
\label{eq:tau_obs}
\bar{\tau}_{\rm obs}(\thetac) = \frac{(X_H+1)\sigma_T}{2m_{\rm amu}}\int \frac{d^2\ell}{(2\pi)^2}U(\ell\thetac)\rho^{\rm prj}_{\rm gal}(\vec{\ell})B(\vec{\ell})\,.
\eeq
Here $U(x)$ is the Fourier transform of the AP filter~\cite{Alonso2016}
\beq
\label{eq:AP_filter}
U(x) = 2\left[2\frac{J_1(x)}{x}-2\frac{J_1(\sqrt{2}x)}{\sqrt{2}x}\right]
\eeq
with $J_1$ being the first Bessel function of the first kind, and $\rho_{\rm gas}^{\rm prj}(\vec{\ell})$ being the Fourier transform of $\rho_{\rm gas}^{\rm prj}(d_A(z)\theta)$. $B(\vec{\ell}) = {\rm e}^{-\sigma_{\rm B}^2\ell^2/2}$ is the Gaussian beam function with $\sigma_{\rm B}={\rm FWHM}/(\sqrt{8} \ln(2))=0.4247 {\rm FWHM}$, and FWHM is the  effective beam full width at half-maximum.

\section{Mock}
\label{sec:mock_obs}

\begin{table}[t!]
\footnotesize
\centerline{\begin{tabular}{cccccccccc}
\hline\hline
 &FWHM & Noise &  Redshift & $V$ & $\bar{n}_{\rm g}$  & $\bar{M}$ & $f_{\rm sky}$\\
&$[{\rm arcmin}]$ & $[\mu K\mathchar`-{\rm arcmin}]$ &  & \ZY{$[(h^{-1}\,{\rm Gpc})^{3}]$} & \ZY{$[({\rm Gpc}^{-1}h)^{3}]$} & $[\msun/h]$  \\
\hline
CMB-S4-like + Next-LSS & $1$ & $2$ & 0.8 $(0.6<z<1.0)$ & 37.8 & $2\times10^{-4}$ & $2.6\times10^{13}$ & 1\\
\hline\hline
\end{tabular}}
\caption{The assumed ideal survey specifications for forecasts. The first two columns show the beam size and detector noise \ZY{of the sky intensity maps} of a CMB-S4-like CMB experiment. The 3rd to 7th column shows the setting of a next generation LSS survey, including the redshift $z$, the comoving survey volume $V$, the mean comoving number density $\bar{n}_{\rm g}$ of clusters, the averaged halo mass $\bar{M}$, and the overlapping sky fraction $f_{\rm sky}$ between the LSS and CMB surveys. Here $\bar{n}_{\rm g}$ and $\bar{M}$  are calculated from the GR part of the ELEPHANT simulation suite~\citep{Cautun2018}. }
\label{table:survey}
\end{table}
\begin{figure}[t!]
\begin{center}
\includegraphics[width=0.6\columnwidth]{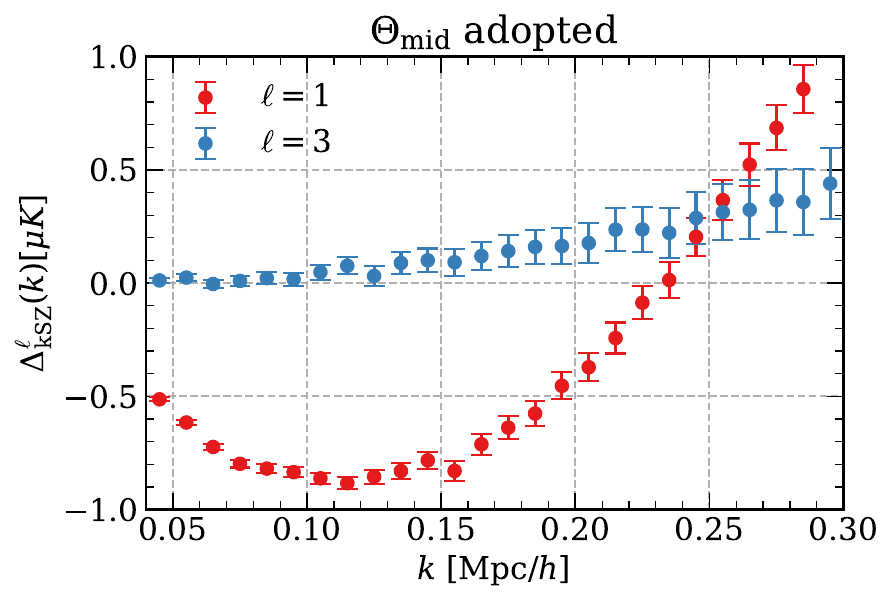}
\end{center}
\caption{Mock pairwise kSZ power spectrum multipoles measured from simulations. $\theta_{\rm c}=2'$ and $\Theta_{\rm mid}$ are adopted, and the assumed survey combination is the Next-LSS+CMB-S4-like survey combination with specifications listed in Table~\ref{table:survey}.}
\label{fig:Pkl1l3}
\end{figure}

In this section we present the mock kSZ observations, including the pairwise kSZ power spectrum in section~\ref{subsec:pkSZ} and the optical depth profiles in section~\ref{subsec:result_tauprof}. We will assume an \ZY{ideal} full-sky survey combination between LSS and CMB experiments to explore the limit of constraints that the next generation surveys can achieve. Then for any real survey setup with $V$ being the overlapping volume between the LSS and CMB surveys, we can practically rescale the constraints by $\sqrt{V_{\rm ideal}/V}$ to obtain  corresponding constraints. This rescaling works at least when we consider scales where the cosmic variance dominates.

\subsection{Mock surveys}
\label{subsec:mock_setup}

Throughout the paper, we assume that we work on a full sky volume-limited galaxy cluster sample, with $M>10^{13}\msun/h$, within the redshift range $0.6<z<1.0$. Such cluster catalogs are expected to be obtained by surveys such as DESI, Euclid and CSST. In this work the mock catalog is constructed from N-body simulations, namely the GR part of the ELEPHANT simulation suite~\citep{Cautun2018}. 
The simulation has a box size of 1024 $\mpch$ and a dark matter particle number of $1024^3$. There are in total 5 independent realizations whose cosmology is GR plus the WMAP9 cosmology. At the snapshot of $z=0.8$, we select all main dark matter halos identified by the group finder \texttt{ROCKSTAR}~\citep{Rockstar} with $M_{\rm 200c}>10^{13}\msun/h$, and these halos of all 5 realizations make a representative sub-sample of our mock cluster catalog. The averages of statistics from this subsample are regarded as those of the whole mock sample. 
Some properties of the mock cluster sample are listed in Table~\ref{table:survey}.

We also assume a full sky CMB-S4-like CMB survey, with an ideal FWHM $\sim 1'$ and a \ZY{detector} noise of 2 $\muKa$~\cite{CMBS42019}. In turn the LSS and CMB surveys are set to have a full-sky celestial overlap. CMB experiments such as AdvACT and SO will have a higher detector noise of~$15\muKa$~\cite{Henderson2016,Simon2019}, which degrades the constraints on the baryonic effect. We will discuss this degradation in section~\ref{subsec:N_det_15}.

\subsection{Mock pairwise kSZ detection}
\label{subsec:pkSZ}

In observations, the density-weighted pairwise kSZ power spectrum can be estimated by~\cite{Sugiyama18}
\bea
P^{\rm s}_{\rm kSZ}(\bfk) &=& \left\langle -\frac{V}{N^2} \sum_{i,j}\left[\delta T_{\rm kSZ}(\hat{n}_i)-\delta T_{\rm kSZ}(\hat{n}_j)\right]e^{-i \bfk\cdot {\bfs}_{ij}} \right\rangle \no \\
&\simeq& \left(\frac{T_0\btau_{\rm obs}}{c}\right)\left\langle \frac{V}{N^2} \sum_{i,j}\left[\bfv_i\cdot\hat{n}_i-\bfv_j\cdot\hat{n}_j\right]e^{-i \bfk\cdot {\bfs}_{ij}} \right\rangle \, \no \\
&\equiv& \left(\frac{T_0\btau_{\rm obs}}{c}\right)P^{\rm s}_{\rm pv}(\bfk)\,,
\label{eq:P_kSZ}
\eea
where $V$ is the survey volume, $N$ is the number of galaxies, ${\bfs}_{ij}={\bfs}_i-{\bfs}_j$ is the pair separation vector in redshift space and $P^{\rm s}_{\rm pv}$ is the density-weighted pairwise LOS velocity power spectrum in redshift space. We have assumed that $\btau$ is not correlated with the galaxy density and velocity fields.

In simulations, $P^{\rm s}_{\rm pv}$ can be calculated via a field based estimator~\cite{Sugiyama2016}
\beq
(2\pi)^3\delta_{\rm D}(\bfk+\bfk')P^{\rm s}_{\rm pv}(\bfk)=\left<p_{\rm s}(\bfk)\delta_{\rm s}(\bfk')-\delta_{\rm s}(\bfk)p_{\rm s}(\bfk')\right> \,,
\eeq
where $p_{\rm s}(\bfk)$ and $\delta_{\rm s}(\bfk)$ are Fourier counterparts of the radial momentum field $p_{\rm s}({\bfs})=[1+\delta_{\rm s}({\bfs})][\bfv({\bfs})\cdot \hat{n}]$ and the density field $\delta_{\rm s}({\bfs})$ respectively. We sample the $p_{\rm s}(\bfs)$ and $\delta_{\rm s}(\bfs)$ fields on $1024^3$ regular grids using the nearest-grid-point (NGP) method and calculate $p_{\rm s}(\bfk)$ and $\delta_{\rm s}(\bfk)$ fields by the fast Fourier transform (FFT).


In figure~\ref{fig:Pkl1l3} we show the mock pairwise kSZ power spectrum multipoles estimated by eq.~(\ref{eq:P_kSZ}), in which $P_{\rm pv}^{\rm s}$ is measured from simulations and $\bar{\tau}_{\rm obs}$ is estimated theoretically as described in section~\ref{sec:theory}. The default BCM parameter set $\Theta_{\rm mid}$ and $\theta_{\rm c} = 2'$ are adopted. The covariance matrix are calculated theoretically based on the measured  momentum-density cross-power spectrum, momentum-momentum auto-power spectrum, and density-density auto-power spectrum in redshift space from simulations~\cite{Sugiyama2017,Zheng20}. We refer readers to the appendix C of~\cite{XiaoL2023} for details of the covariance matrix estimation, where the accuracy of the theoretical covariance matrix is validated against simulation based estimation. It was shown there that the off-diagonal components of the covariance can be ignored and the accuracy of the theoretical covariance matrix  will not bias the conclusion of this work.





\subsection{Mock $\btau_{\rm obs}$ profile measurements}
\label{subsec:result_tauprof}
\begin{figure}[t!]
\begin{center}
\includegraphics[width=0.54\columnwidth]{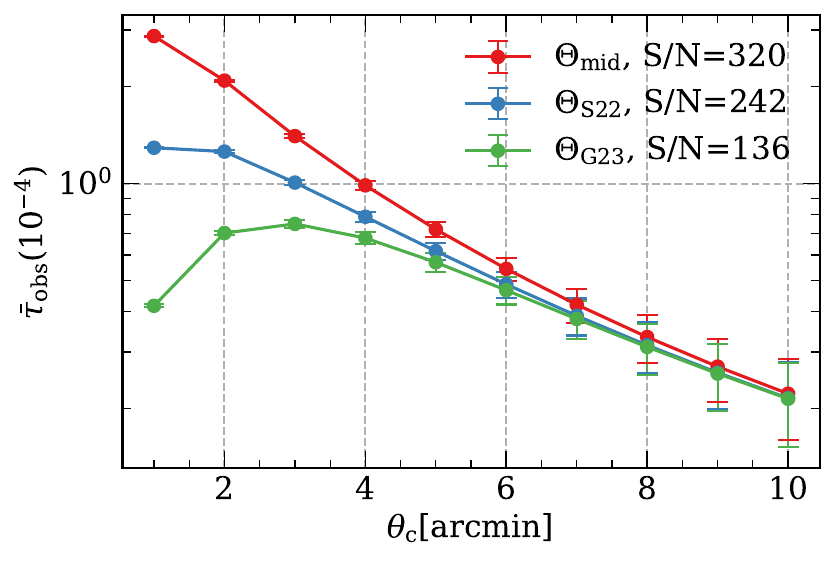}
\includegraphics[width=0.42\columnwidth]{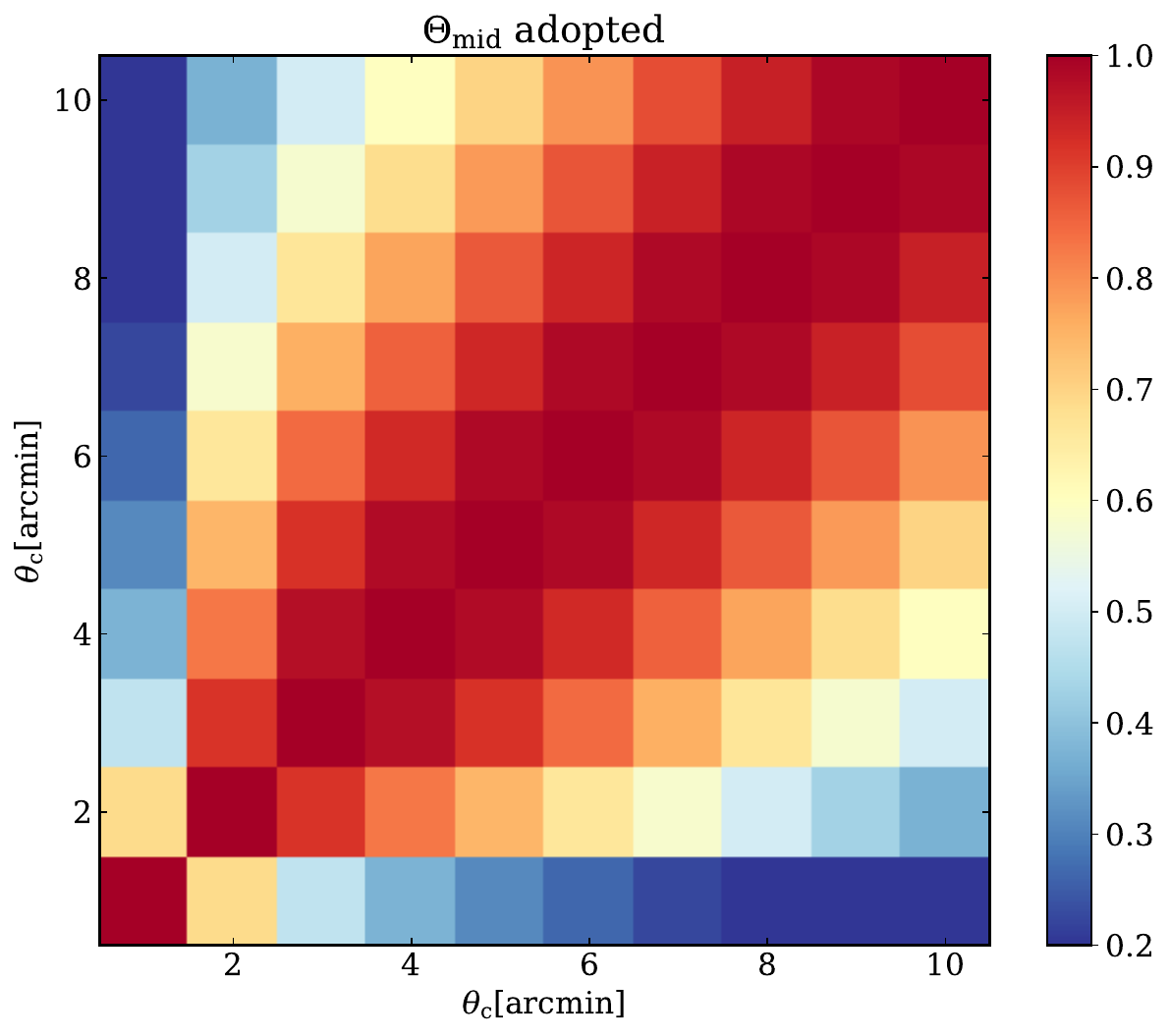}
\end{center}
\caption{\textit{Left panel:} mock $\btau_{\rm obs}(\thetac)$ profiles \ZY{(eq.~(\ref{eq:tau_obs}))} with error bars expected to be measured with different sets of BCM parameter values in table~\ref{table:para}. \textit{Right panel:} the correlation coefficient matrix when the default BCM parameters $\Theta_{\rm mid}$ are adopted.}
\label{fig:tau_profile}
\end{figure}
Assuming that we have a perfect model for $P^{\rm s}_{\rm pv}$ coming from e.g. the RSD analysis~\cite{Sugiyama2017}, we solve for the constraint on $\btau_{\rm obs}$ by eq.~(\ref{eq:P_kSZ}). In observations, we can vary $\thetac$, measure the corresponding $P_{\rm kSZ}$, and then obtain a measurement of the $\btau_{\rm obs}$ profile~\cite{Sugiyama18}. It is this profile that encodes information of the baryon density distribution within halos~\cite{Amodeo2021} and we will study its constraining power on  BCM parameters in section~\ref{sec:results}.

On the left panel of figure~\ref{fig:tau_profile}, we set $\thetac=1',2',...9',10'$ and calculate the $\btau_{\rm obs}(\thetac)$ following section~\ref{sec:theory}. Different fiducial $\Theta$ values yield different $\btau_{\rm obs}$ profiles. As shown in figure~\ref{fig:tau_sensitivity} of appendix~\ref{app:tau_BCM_dependency}, the amplitude of the $\btau_{\rm obs}$ is most sensitive to $\log_{10} M_{\rm c}$, in particular at inner regions of halos. A larger $\log_{10} M_{\rm c}$ induces a larger $\beta$, a flatter $\rho_{\rm gas}$ and in turn a lower central $\btau_{\rm obs}$ after the CMB beam smoothing. As a result, $\Theta_{\rm G23}$, having the largest $\log_{10} M_{\rm c}$, yields the lowest and most flattened $\btau_{\rm obs}$ profile within the three.

The covariance matrix ${\rm Cov}(\theta_{{\rm c},i},\theta_{{\rm c},j})$, whose diagonal elements being the error bars of $\btau_{\rm obs}$ profiles, is evaluated following eq.~(65) of~\cite{Sugiyama18}, where we use $P_{\rm kSZ}$ measurements up to $k=0.3\hmpc$ during the $\btau_{\rm obs}$ fitting. On the right panel of figure~\ref{fig:tau_profile}, the correlation coefficient matrix ${\rm \widetilde{Cov}}(\theta_{{\rm c},i},\theta_{{\rm c},j})\equiv {\rm Cov}(\theta_{{\rm c},i},\theta_{{\rm c},j})/\sqrt{{\rm Cov}(\theta_{{\rm c},i},\theta_{{\rm c},i}){\rm Cov}(\theta_{{\rm c},j},\theta_{{\rm c},j})}$ with the default $\Theta_{\rm mid}$ is displayed. The $\btau_{\rm obs}$ error bar increases towards large $\thetac$. 
Moreover, the correlation between adjacent $\thetac$ bins are higher for larger $\thetac$ bins by construction, due to the fact that larger adjacent $\thetac$ bins share more overlapping regions. As a result, most S/N of the $\btau_{\rm obs}$ profile observation is contributed by small $\thetac$'s and by inner regions of halos, which demands a high resolution of the CMB experiment when detecting the kSZ effect. {This fact also indicates an expectation that the kSZ effect will most tightly constrain $\log_{10} M_{\rm c}$, which is most sensitive to the inner halo region gas density, as demonstrated in section~\ref{sec:results}.}

The S/N of the $\btau$ profile measurement can be estimated by
\beq
\left(\frac{S}{N}\right)^2 = \sum_{i,j} \btau_{\rm obs}(\theta_{{\rm c},i}){\rm Cov}^{-1}(\theta_{{\rm c},i},\theta_{{\rm c},j})\btau_{\rm obs}(\theta_{{\rm c},j})\,. 
\label{eq:tau_prof_SN}
\eeq
For our ideal survey combination, it reaches $\rm S/N_{\rm ideal}$=320/242/136 when $\Theta_{\rm mid}$/$\Theta_{\rm S22}$/$\Theta_{\rm G23}$ is adopted. As expected, the detection S/N depends on the chosen fiducial BCM parameter values. A lower $\bar{\tau}_{\rm obs}$ profile, in particular at small $\thetac$ bins, corresponds to a lower S/N. Despite this, all three S/N's reach $\mathcal{O}(100)$, 2 orders of magnitude higher than that of the current kSZ detections (e.g.,~\cite{PlanckkSZ18,Schaan2021,Chen2022}). For a real observation with a smaller overlapping survey volume $V$ at $z\sim0.8$, the S/N can be roughly estimated by ${\rm S/N}_{\rm ideal}\sqrt{V/37.8{\rm Gpc}^3h^{-3}}$.

\section{Results}
\label{sec:results}

In this section, we apply the fisher matrix technique and predict the constraints that future surveys will impose on BCM parameters. For a Gaussian likelihood function, the Fisher matrix can be calculated by
\beq
\label{eq:fisher_m}
F_{ij} = \frac{\partial X^{\rm T}}{\partial \theta_i}{\rm Cov}(x_i,x_j)^{-1}\frac{\partial X}{\partial \theta_j}+(prior)\,.
\eeq 
Here we have the vector of model parameters $\Theta \equiv \{\theta_i\}$, and $X=X(\Theta)\equiv\{x_i\}$ is the vector of the measured quantities. ${\rm Cov}(x_i,x_j)$ is the data covariance matrix which is assumed to be independent of $\Theta$. The \textit{prior} term is usually a diagonal matrix with its diagonal element being $1/\sigma_i^2$, in which $\sigma_i^2$ is the variance of the Gaussian prior of the $i$th parameter. We apply flat priors $[a_i,a_i+b_i]$ in this work, and in practice we simply regard $(b_i/2)^2$ as the variance of the equivalent Gaussian prior. 

In an ideal case of the Gaussian likelihood, the covariance matrix between parameters ${\rm Cov}(\theta_i,\theta_j)=F^{-1}$, and $\Delta \theta_i\ge(F^{-1})^{1/2}_{ii}$ according to the Cram\'er–Rao inequality. Therefore we can place a firm lower limit on the parameter error bar $\Delta \theta_i$ that can be attained from surveys. We will quote this lower limit as the forecast of the parameter constraint in this work.

\subsection{Constraints on BCM parameters}
\label{subsec:cons_BCM}
\begin{figure}[t!]
\begin{center}
\includegraphics[width=0.52\columnwidth]{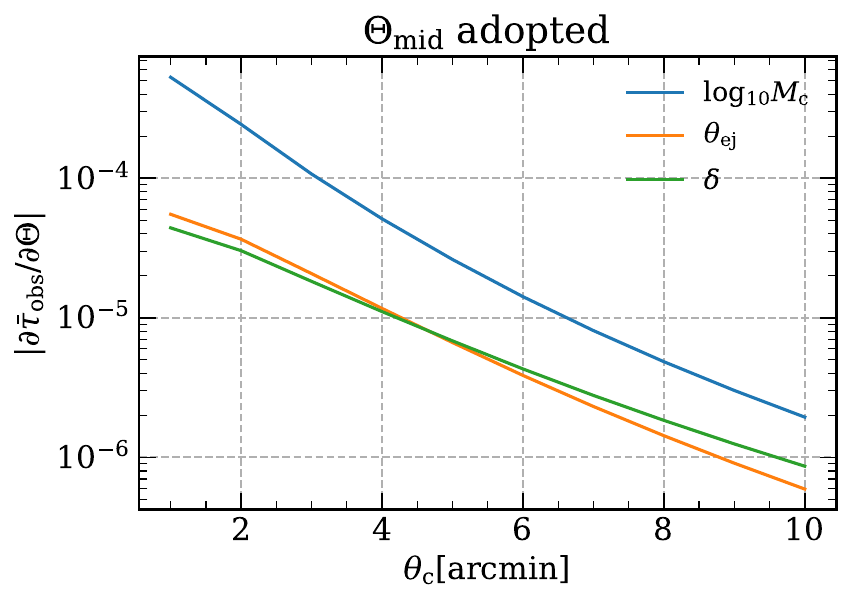}
\includegraphics[width=0.44\columnwidth]{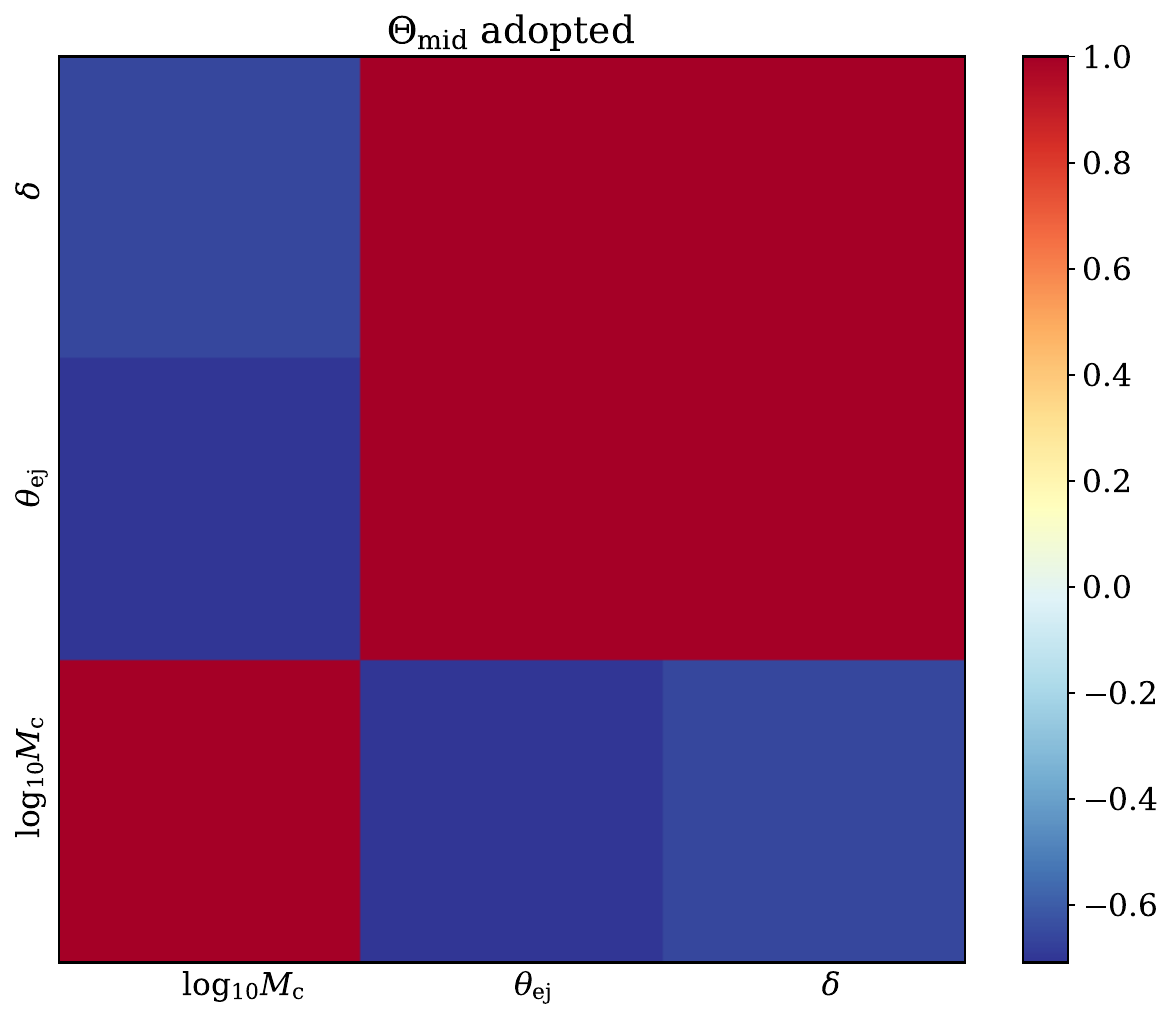}
\end{center}
\caption{\textit{Left panel:} absolute values of the derivatives of $\btau_{\rm obs}$ with respect to 3 kSZ-sensitive BCM parameters in table~\ref{table:para}. \textit{Right panel:} the correlation coefficient matrix between 3 kSZ-sensitive BCM parameters. Only the case with $\Theta_{\rm mid}$ is shown.}
\label{fig:deri_3para}
\end{figure}
\begin{figure*}[t!]
\begin{center}
\includegraphics[width=0.8\columnwidth]{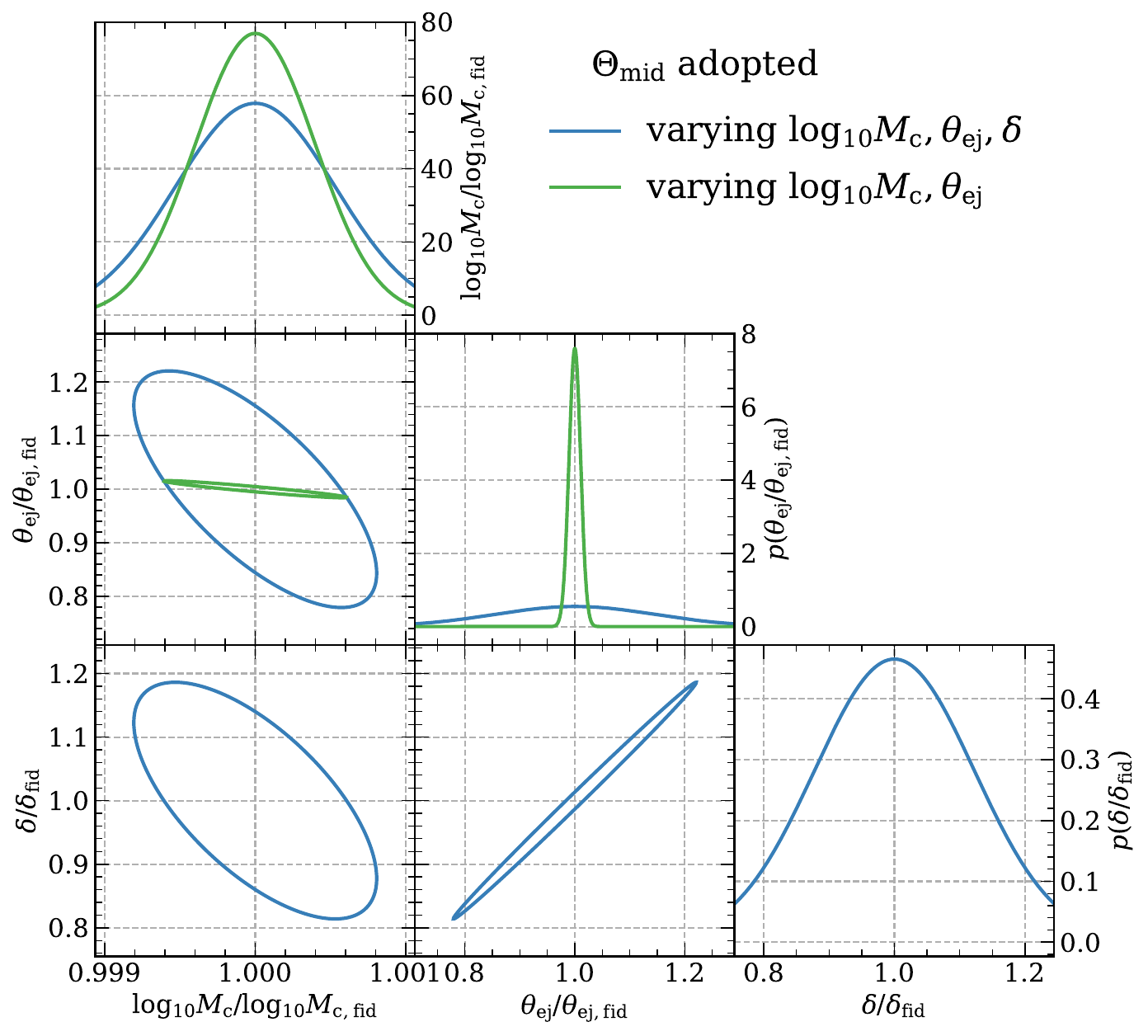}
\end{center}
\caption{ Predicted 1-$\sigma$ contours of the constraints on 3 kSZ-sensitive BCM parameters when the default $\Theta_{\rm mid}$ is adopted. Blue lines show the case when we vary $\log_{10} M_{\rm c}$, $\theta_{\rm ej}$, $\delta$. Green lines show the case when we further fix $\delta$ and  vary $\log_{10} M_{\rm c}$ and $\theta_{\rm ej}$. On diagonal panels we show the corresponding 1-D marginalized parameter posterior PDF.}
\label{fig:BCM_contours}
\end{figure*}

\begin{table}[t!]
\centering
\begin{tabular}{ccccc}
\hline\hline
 & $\Delta\log_{10} M_{\rm c}/\log_{10} M_{\rm c}$ & $\Delta\theta_{\rm ej}/\theta_{\rm ej}$ & $\Delta\delta/\delta$ &  \\
\hline
$\log_{10} M_{\rm c},\theta_{\rm ej},\delta$ & $0.05\%$ & $14.52\%$ & $12.25\%$  \\
$\log_{10} M_{\rm c},\theta_{\rm ej}$ & $0.04\%$ & $1.05\%$ & - \\
Only 1 para & $0.01\%$ & $0.32\%$ & $0.29\%$ \\
\hline\hline
\end{tabular}
\caption{1-$\sigma$ error  of BCM parameter constraints, when $\Theta_{\rm mid}$ is adopted as fiducial values. As an exhibition of the best constraints we can obtain, in the last row we show the individual constraint on each parameter when we only vary it and fix other ones.}
\label{table:para_error}
\end{table}
{Before applying the Fisher matrix technique, it is useful to visually check the degeneracy between parameters by comparing the derivatives of $\btau_{\rm obs}$ with respect to BCM parameters. These $|\partial \btau_{\rm obs}/\partial \Theta|$'s are shown on the left panel of figure~\ref{fig:deri_3para}. As we can see, the amplitude of $|\partial \btau_{\rm obs}/\partial \log_{10} M_{\rm c}|$ is one order of magnitude higher than those of $|\partial \btau_{\rm obs}/\partial \theta_{\rm ej}|$ and $|\partial \btau_{\rm obs}/\partial \delta|$, and the latter two derivative lines nearly overlap with other. We thus expect that the kSZ detection will tightly constrain $\log_{10} M_{\rm c}$, while constraints on $\theta_{\rm ej}$ and $\delta$ will be loose due to the their high degeneracy.  This visual impression matches  the correlation coefficient matrix on the right panel of figure~\ref{fig:deri_3para}. Indeed, the correlation coefficient between $\theta_{\rm ej}$ and $\delta$ is nearly 1, which is expected as they both characterize the gas density profile at outer region of halos. 

We display 1-$\sigma$ contours of  parameter posterior distribution in figure~\ref{fig:BCM_contours}. As expected, the constraint on $\log_{10} M_{\rm c}$ is tight, while the contour between $\theta_{\rm ej}$ and $\delta$ is heavily elongated. It is natural to expect that, if extra data such as X-ray detection and tSZ observation can help break the $\theta_{\rm ej}-\delta$ degeneracy, constraints on both parameters will be heavily improved. We show this case by blue contours in figure~\ref{fig:BCM_contours}, where we fix $\delta$ to its fiducial value and only set free $\log_{10} M_{\rm c}$ and $\theta_{\rm ej}$ in the Fisher matrix calculation. The constraint of $\theta_{\rm ej}$ is improved by one order of magnitude. As a reference, in table~\ref{table:para_error} we list the 1-$\sigma$ constraints of BCM parameters when $\Theta_{\rm mid}$ is adopted as fiducial values.}

\subsection{Constraints on the matter power spectrum damping}
\label{subsec:cons_D_BCM}
\begin{figure}[t!]
\includegraphics[width=1\columnwidth]{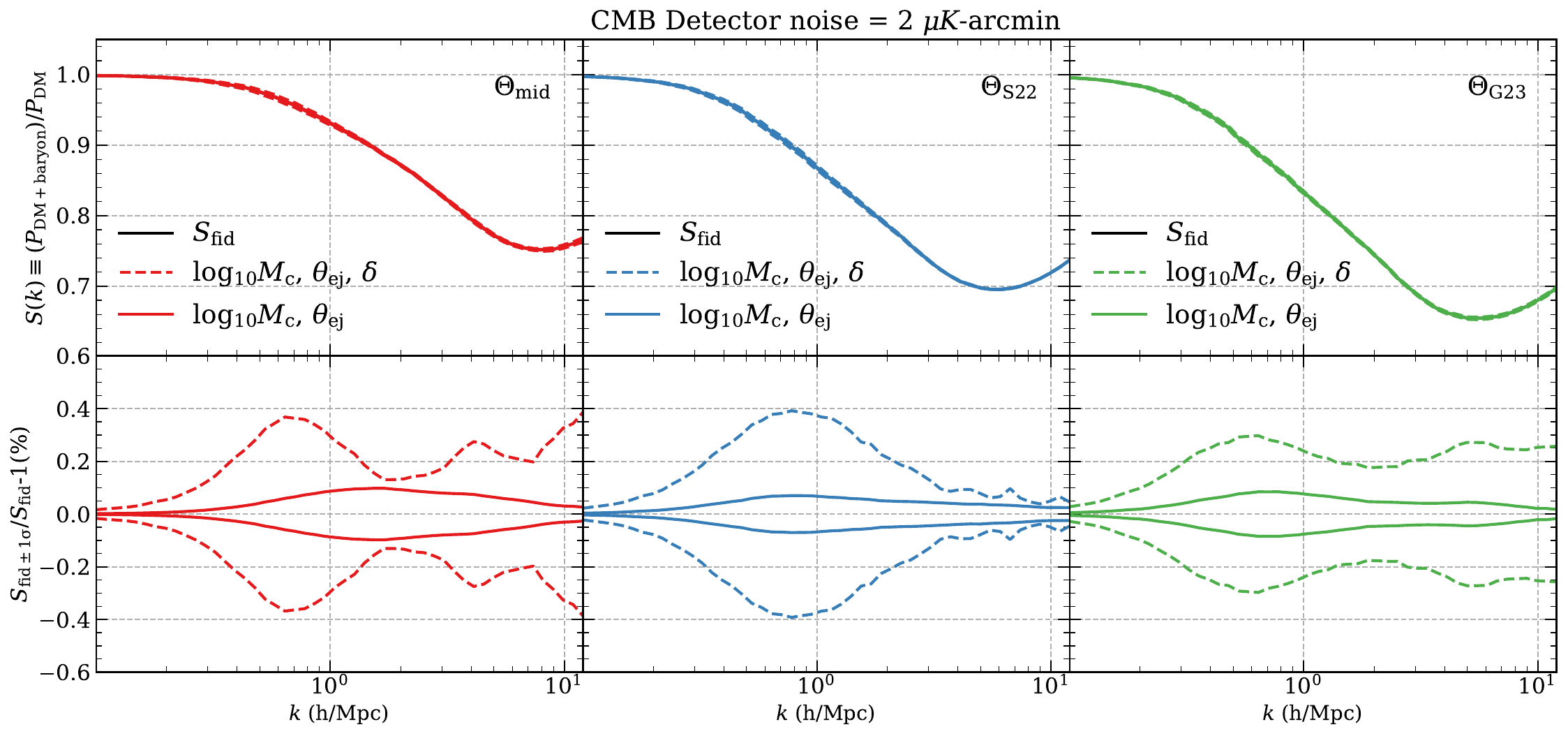}
\caption{The predicted 1-$\sigma$ variations of $S(k)$ when different fiducal BCM parameter values are adopted. Here the CMB experiment is assumed to have a detector noise of $2\mu K\mathchar`-{\rm arcmin}$.}
\label{fig:cons_D_BCM_multipara_2muK}
\end{figure}

\ZY{Now we turn to propagate uncertainties of BCM parameters to constraints of the matter power spectrum damping $S(k)\equiv P_{\rm DM+baryon}/P_{\rm DM}$, where $P_{\rm DM+baryon}$ is the matter power spectrum of the dark matter+baryon fluid (with baryonic feedback), and $P_{\rm DM}$ is the matter power spectrum when the matters is only composed of dark matter (without baryonic feedback)}. We adopt the \texttt{BCemu} emulator developed in~\cite{Giri2021} to map BCM parameters to $S(k)$\footnote{There is a 7th parameter $\delta\eta$ as an input in the \texttt{BCemu} emulator. We follow the well constrained relation $\eta_{\rm star}(\delta\eta/0.24)^{0.19}=0.239$ in~\cite{Grandis2023} to infer the $\delta\eta$ value from $\eta_{\rm star}$.}. As illustrated in appendix~\ref{app:sensitivity_D_BCM},  BCM parameters are also degenerated with each other in contributing $S(k)$. In order to tackle this degeneracy, the 1-$\sigma$ variation of $S(k)$ can be evaluated by
\beq
\sigma_{S}^2(k) = \sum_{ij} S_{,i}S_{,j}(F^{-1})_{ij} + \frac{1}{2}\sum_{ij}\sum_{lm}S_{,ij}S_{,lm}(F^{-1})_{il}(F^{-1})_{jm}\,,
\label{eq:S_fisher}
\eeq
where $F$ is the Fisher matrix between BCM parameters,  $S_{,i}$ is the first-order partial derivative of $S(k)$ with respect to the $i$th BCM parameter, and $S_{,ij}$ is the second-order partial derivative of $S(k)$ with respect to the $i$th and $j$th BCM parameters. The derivation and validation of eq.~(\ref{eq:S_fisher}) is presented in appendix~\ref{app:S_fisher}, and the 1-$\sigma$ constraints on $S(k)$, $\sigma_S(k)$, are shown in figure~\ref{fig:cons_D_BCM_multipara_2muK}.

Next generation of LSS analysis will achieve a $1\%$ precision on constraining cosmological parameters. It then requires stronger controls of all systematic errors. In the WL case, one crucial requirement is that $\sigma_S(k)<1\%$ at $k\lesssim 1-10\hmpc$. As shown in figure~\ref{fig:cons_D_BCM_multipara_2muK}, in the 3\ZY{-params} case of varying $\log_{10} M_{\rm c}$, $\theta_{\rm ej}$ and $\delta$ in the analysis,  $\sigma_S(k)$ can be constrained to a level of $\sigma_S(k)<0.4\%\sqrt{37.8{\rm Gpc}^3h^{-3}/V}$ at $k\lesssim10\hmpc$. When we further fix $\delta$ and only vary $\log_{10} M_{\rm c}$ and $\theta_{\rm ej}$ (2\ZY{-params} case), $\sigma_S(k)$ decreases to $\sigma_S(k)<0.1\%\sqrt{37.8{\rm Gpc}^3h^{-3}/V}$. Here we have applied the scaling factor $\sqrt{37.8{\rm Gpc}^3h^{-3}/V}$ so as to have a rough idea of what is going on for a realistic survey combination in the future. 

As an example, DESI, Euclid and CSST  will likely have an overlapping area of $8000 {\rm deg}^2$ with CMB-S4, then at $0.6<z<1.0$ and $k\lesssim10\hmpc$, we can obtain $\sigma_S(k)<0.9\%$ for the 3\ZY{-params} fitting case, and $\sigma_S(k)<0.22\%$ at $k\lesssim10\hmpc$ for the 2\ZY{-params} fitting case.  For LSST which will likely have a $18000 {\rm deg}^2$ overlapping area, we can achieve $\sigma_S(k)<0.6\%$ and $\sigma_S(k)<0.15\%$ for the 3- and 2\ZY{-params} fitting cases. We list these numbers in table~\ref{table:example}. 

\ZY{Moreover, figure~\ref{fig:cons_D_BCM_multipara_2muK} plots three $\sigma_S(k)$'s when three $\Theta$'s listed in table~\ref{table:para} are adopted. We notice that, conclusions above are generally immune from the variation of the fiducial BCM parameter values, while subtle differences do exist. For example, at small scales, the 3-params constraint on $S(k)$ when $\Theta_{\rm S22}$ is adopted is decreasing, which is opposite to cases of $\Theta_{\rm mid}$ and $\Theta_{\rm G23}$. This trend depends on  different fiducial values adopted, such as the $\mu = 0.3$ of $\Theta_{\rm S22}$ and $\mu = 1$ of $\Theta_{\rm mid}$, through complicated degeneracies between parameters induced when they determine $\bar{\tau}_{\rm obs}$ and $S(k)$. This is confirmed in appendix~\ref{app:mu_dependency}, where we show how the variation of $\mu$ value changes this scale dependent constraining trend.}

\ZY{Although the appendix~\ref{app:mu_dependency} shows only the $\mu$ dependence of the trend, we emphasize here that this trend does depend on all BCM parameters in a complicated way. One evidence for this is that $\Theta_{\rm mid}$ and $\Theta_{\rm G23}$ have similar scale dependence of the $S(k)$ constraint when they are qualitatively very different from each other. A thorough study of this $\Theta$ value dependency needs considering more variations in the parameter space, which is left to the future work.}

\subsection{CMB experiments with a detector noise of $15\mu K\mathchar`-{\rm arcmin}$}
\label{subsec:N_det_15}
\begin{figure}[t!]
\includegraphics[width=1\columnwidth]{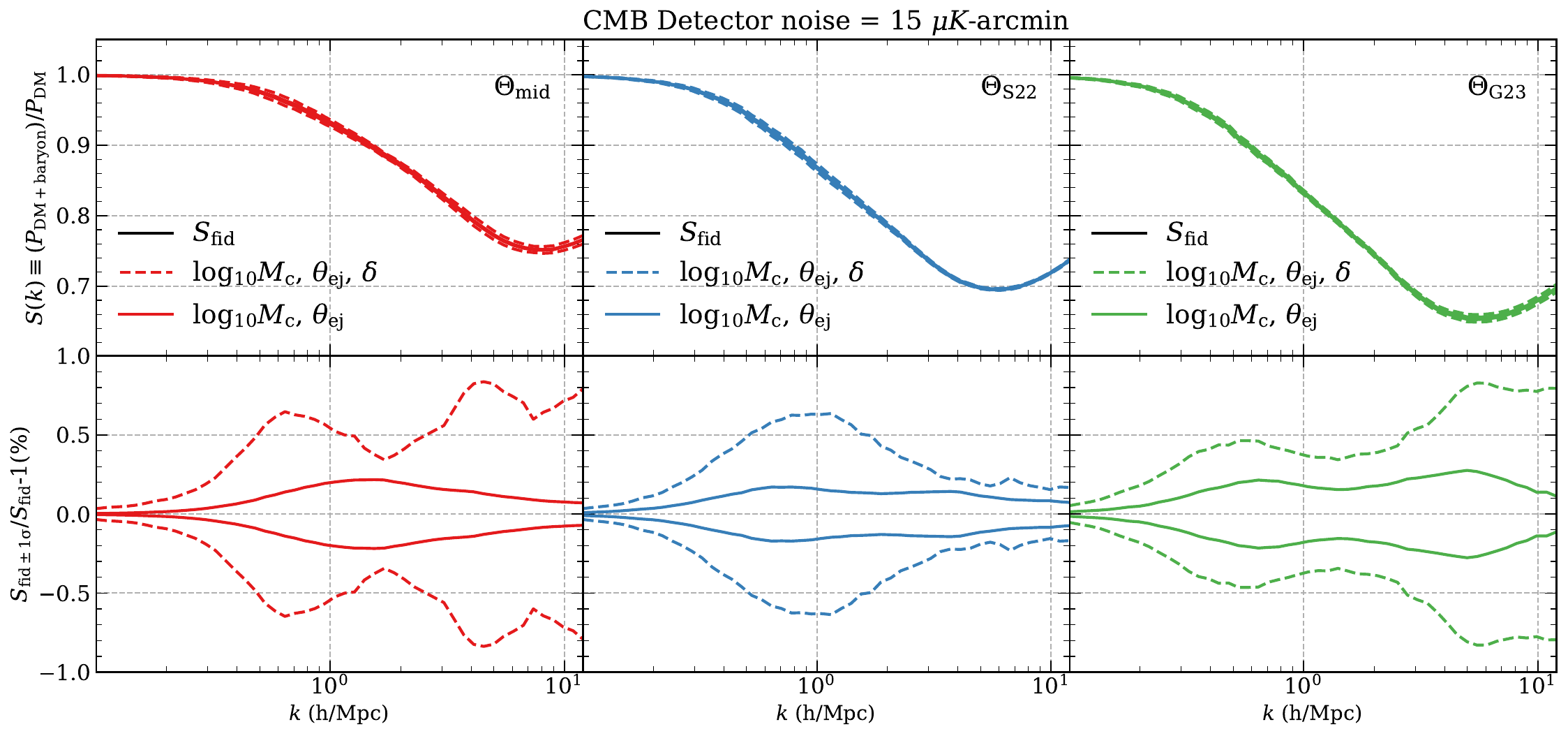}
\caption{Same as figure~\ref{fig:cons_D_BCM_multipara_2muK}, except that the CMB experiment here is assumed to have a detector noise of $15\mu K\mathchar`-{\rm arcmin}$.}
\label{fig:cons_D_BCM_multipara_15muK}
\end{figure}
\begin{table}[t!]
\centering
\begin{tabular}{c|cccc}
\hline\hline
 & DESI/Euclid/CSST & LSST &  \\
\hline
CMB-S4/CMB-HD & $0.9\%$($0.22\%$) & $0.6\%$($0.15\%$)  \\
AdvACT/SO & $1.8\%$($0.7\%$) & $1.2\%$($0.5\%$)  \\
\hline\hline
\end{tabular}
\caption{{1-$\sigma$ constraints on $S(k)$ at $k\lesssim10\hmpc$ of some future survey combinations in the 3\ZY{-params} (2\ZY{-params}) fitting case.}}
\label{table:example}
\end{table}
{While CMB-S4 and CMB-HD aim to reduce the detector noise down to $2\muKa$ or lower~\cite{CMBS42019,CMBHD2020}, experiments such as AdvACT and SO will have a higher noise at ~$15\muKa$~\cite{Henderson2016,Simon2019}. This high detector noise will enhance the noise of the kSZ power spectrum at small scales and in turn reduce the S/N of the measured $\bar{\tau}_{\rm obs}$ profile. In order to test its impact on the $\sigma_S(k)$ constraint, we repeat the calculations above except that we set the detector noise to $15\muKa$ now. The results are shown in figure~\ref{fig:cons_D_BCM_multipara_15muK}.

As expected, the 1-$\sigma$ constraint on $S(k)$ becomes looser. The resultant $\sigma_S(k)$ looks doubled or tripled  compared to that of the $2\muKa$ detector noise case.  In the 3\ZY{-params} fitting case, we have $\sigma_S(k)<0.8\%\sqrt{37.8{\rm Gpc}^3h^{-3}/V}$ at $k\lesssim10\hmpc$, and in the 2\ZY{-params} fitting case, we have $\sigma_S(k)<0.3\%\sqrt{37.8{\rm Gpc}^3h^{-3}/V}$ at $k\lesssim10\hmpc$. Some examples of predictions are listed in table~\ref{table:example}.} 

\section{Conclusion}
\label{sec:conclusion}

The baryonic feedback effect is an important systematic error in the weak lensing (WL) analysis. It prevents the small scale WL data from being used to constrain cosmology and contribute partly to the $S_8$ tension. With the coming generations of  large scale structure (LSS) and CMB experiments, the S/N of the kSZ detection will be improved to a level of $\mathcal{0}(100)$. This high precision of the kSZ  detection will tightly constrain the baryon distribution in and around dark matter halos, which (1) helps us understand physical mechanisms driving dark matter-baryon co-evolution at halo scales, and (2) quantify the baryonic effect in the weak lensing statistics. 

In this work, we apply the Fisher matrix technique to analyze the kSZ constraints on three kSZ-sensitive Baryon Correction Model (BCM) parameters ($\log_{10} M_{\rm c}$, $\theta_{\rm ej}$ and $\delta$). Our calculations show that, in combination with next generation LSS surveys, the 3rd generation CMB experiments such as AdvACT and Simon Observatory can constrain the matter power spectrum damping $S(k)$ to the precision of $\sigma_S(k)<0.8\%\sqrt{37.8{\rm Gpc}^3h^{-3}/V}$ at $k\lesssim 10\hmpc$, where $V$ is the overlapped survey volume between the future LSS and CMB surveys. For the 4th generation CMB surveys such as CMB-S4 and CMB-HD, the constraint will be enhanced to $\sigma_S(k)<0.4\%\sqrt{37.8{\rm Gpc}^3h^{-3}/V}$ at $k\lesssim 10\hmpc$. If extra-observations, e.g. X-ray detection and tSZ observation, can constrain and effectively fix the gas density profile slope parameter $\delta$, the constraint on $S(k)$ will be further boosted to $\sigma_S(k)<0.3\%\sqrt{37.8{\rm Gpc}^3h^{-3}/V}$ at $k\lesssim 10\hmpc$ and $\sigma_S(k)<0.1\%\sqrt{37.8{\rm Gpc}^3h^{-3}/V}$ at $k\lesssim 10\hmpc$ for the 3rd and 4th generation CMB surveys.

One major simplification in our analysis is that we only vary three kSZ-sensitive parameters in the Fisher matrix analysis. Other BCM parameters are assumed to be constrained and fixed by extra observations such as X-ray, tSZ and optical data etc.~\cite{Schneider2020b}. A thorough investigation of how BCM and $S(k)$ can be constrained by a joint future data analysis will be left to future works.

\acknowledgments
YZ acknowledges the supports from the National Natural Science Foundation of China (NFSC) through grant 12203107, the Guangdong Basic and Applied Basic Research Foundation with No.2019A1515111098, and the science research grants from the China Manned Space Project with NO.CMS-CSST-2021-A02. PJZ acknowledges the supports from  the National Key R\&D Program of China (2023YFA1607800, 2023YFA1607801).

\appendix
\section{BCM parameter dependence of the $\btau$ profile}
\label{app:tau_BCM_dependency}
\begin{figure*}[t!]
\begin{center}
\includegraphics[width=\columnwidth]{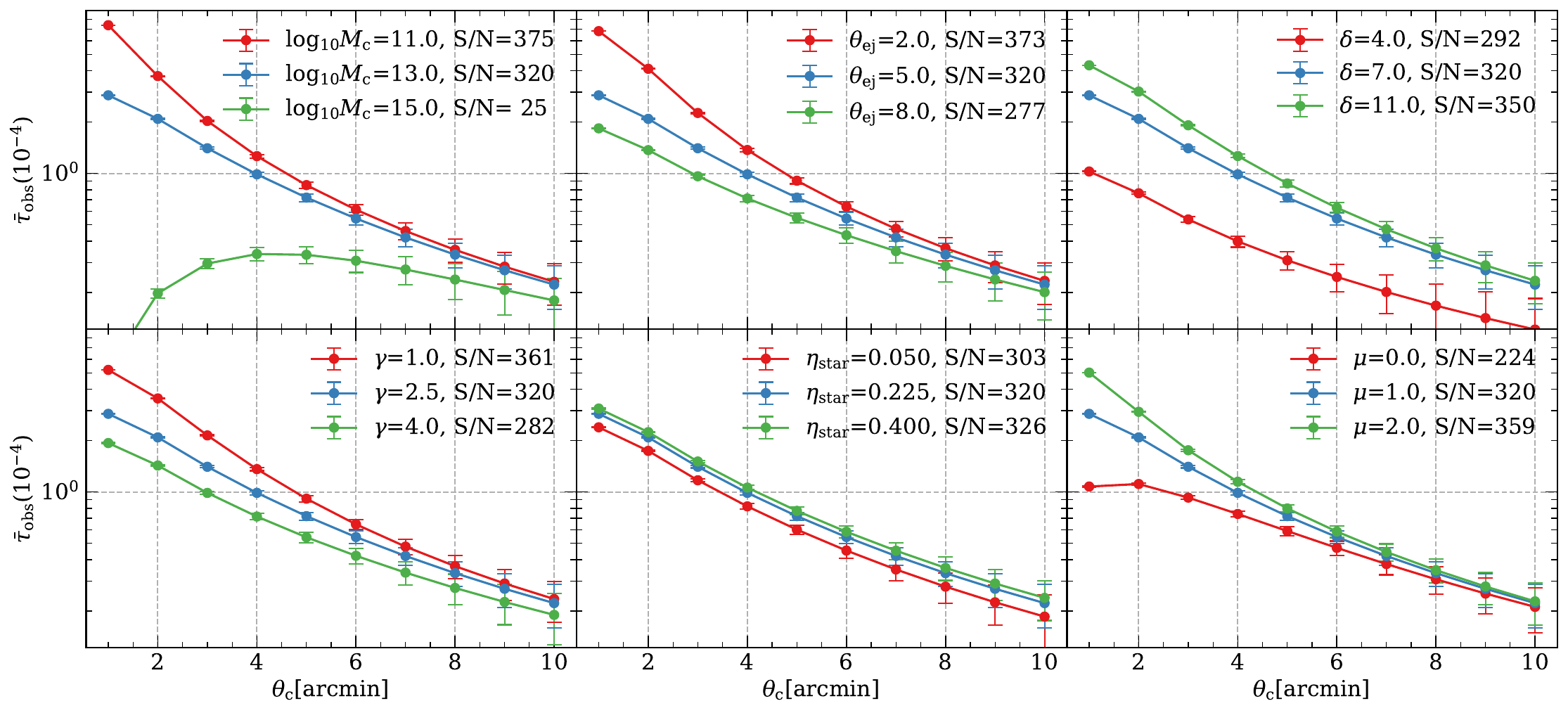}
\end{center}
\caption{The sensitivity test of the mock $\btau_{\rm obs}$ profile with respect to 6 BCM parameters. In each panel, we vary only one BCM parameter and fix the other five to their $\Theta_{\rm mid}$ values. \ZY{On the top left panel, a non-physical case of eq.~(\ref{eq:rho_gas0}) being divergent when $\delta=3$ is discarded. Instead we show the result when $\delta=4$.}}
\label{fig:tau_sensitivity}
\end{figure*}
In this appendix, we test the sensitivity of the mock $\btau_{\rm obs}$ profile on the BCM parameter variation. We change only one parameter at one time and present the corresponding $\btau_{\rm obs}$ profiles on each panel of figure~\ref{fig:tau_sensitivity}. \ZY{The variation ranges fulfill the prior ranges in table~\ref{table:para}\footnote{\ZY{One exception is the $\delta$ parameter. Its lower bound of the prior range ($\delta=3$) leads to a integral divergence of eq.~(\ref{eq:tau_obs}), instead of which we show the case of $\delta=4$ on the top left panel of figure~\ref{fig:tau_sensitivity}.}}}. 


These variations can be uniquely understood by the following argument. \ZY{In the \texttt{S19} model, the total gas fraction of a halo, whose boundary is extended to infinity theoretically, is fixed to  $f_{\rm gas} = \Omega_{\rm b}/\Omega_{\rm m} - f_{\rm star}$. However, observationally,  the AP filter actually measures the integrated baryon content within a fixed region of the halo. Therefore, for a fixed halo mass and a fixed $\thetac$, a larger $f_{\rm star}$ and a flatter gas density profile means that a smaller amount of baryon will be detected by the kSZ effect, denoting a lower $\btau_{\rm obs}$. }

As a result, on the lower middle panel of figure~\ref{fig:tau_sensitivity}, by increasing $\eta_{\rm star}$, we increase $f_{\rm star}$ and decrease $f_{\rm gas}$, so the $\btau_{\rm obs}$ profile moves downward consistently. On other panels, by enhancing $\log_{10} M_{\rm c}$, $\theta_{\rm ej}$ and $\gamma$, and by reducing $\delta$ and $\mu$, we get a flatter gas density profile and in turn a lower and flatter $\btau_{\rm obs}$ profile.

\ZY{Besides, in figure~\ref{fig:tau_sensitivity}, by comparing the top left and bottom right panels, we see that $M_{\rm c}$ is dominant on the $\beta$ evaluation; by comparing the top left and bottom left panels, we find that $\bar{\tau}_{\rm obs}$ is more sensitive to $\beta$ than $\gamma$ in the inner halo regions; and by comparing the top right and bottom left panels, we confirm that $\bar{\tau}_{\rm obs}$ is more sensitive to $\delta$ than $\gamma$ at the outer halo regions. While supporting our free parameter choice in this work, these results also denote considerable degeneracy between $\gamma$, $\mu$ and other parameters. Therefore, by fixing $\gamma$ and $\mu$ in our Fisher analysis, we are effectively considering a joint analysis between the kSZ observation and extra datasets, such as optical data, tSZ and X-ray observations to break the degeneracy and pin down  uncertainties of these parameters.}

\section{1-$\sigma$ variation of $S(k)$ calculation}
\label{app:S_fisher}
\begin{figure*}[t!]
\begin{center}
\includegraphics[width=0.5\columnwidth]{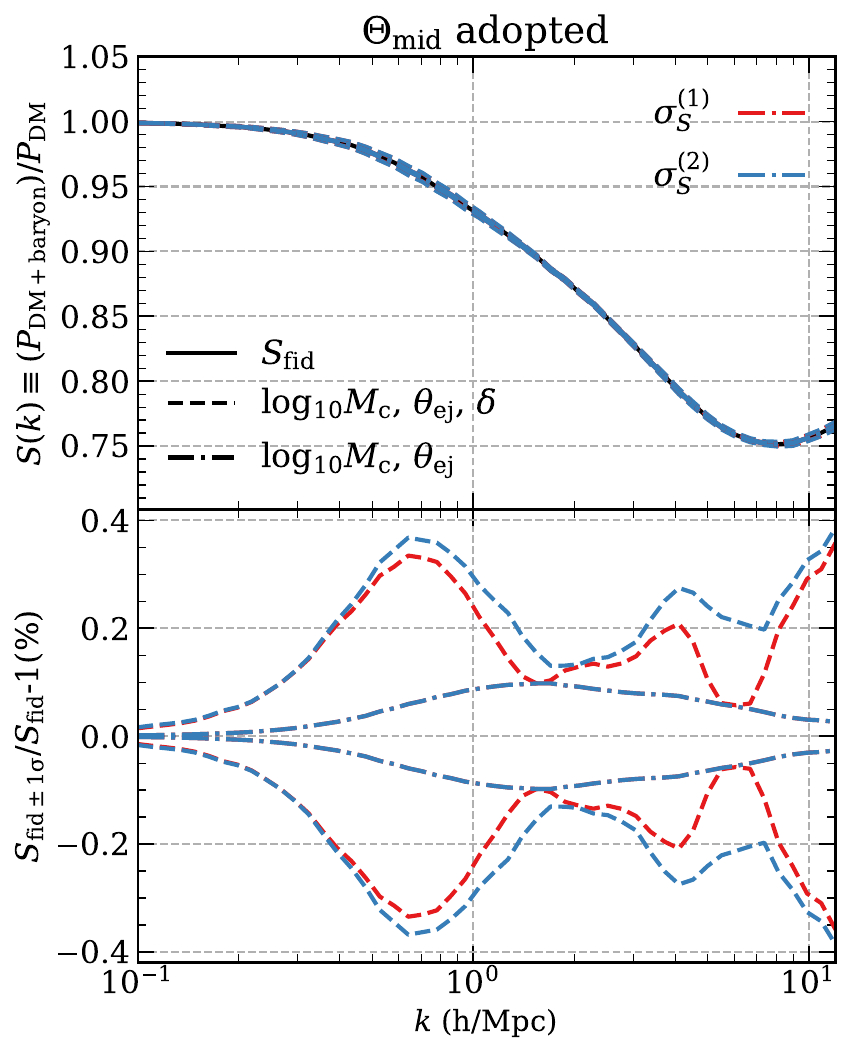}
\end{center}
\caption{Comparison of the predicted 1-$\sigma$ variation of $S(k)$ when we truncate the Taylor expansion of $S(\Theta)$ at the first order of $\delta\theta$, $\sigma^{(1)}_S$ (red lines), and at the second order of $\delta\theta$, $\sigma^{(2)}_S$ (blue lines). The black solid line on the top panel shows the fiducial $S(k)$ function. The dashed and dot-dashed lines show the 1-$\sigma$ variations of $S(k)$ when ($\log_{10}M_{\rm c}$, $\theta_{\rm ej}$, $\delta$) are set to be free parameters or ($\log_{10}M_{\rm c}$, $\theta_{\rm ej}$) are set to be free.}
\label{fig:sigma_S_compare}
\end{figure*}
We Taylor expand $S(k)$ at fiducial values of BCM parameters $\bar{\Theta}$, omitting the $k$ dependence, such as
\beq
\label{eq:S_taylor}
S(\Theta) = S(\bar{\Theta})+\sum_iS_{,i}\delta\theta_i + \frac{1}{2}\sum_{ij}S_{,ij}\delta\theta_i\delta\theta_j\,,
\eeq
Here we truncate at the second order of the expansion, $S_{,i}$ is the first-order partial derivative of $S(k)$ with respect to the $i$th BCM parameter, and $S_{,ij}$ is the second-order partial derivative of $S(k)$ with respect to the $i$th and $j$th BCM parameters.

The mean of $S(\Theta)$ is then
\beq
\left<S(\Theta)\right> = S(\bar{\Theta})+\frac{1}{2}\sum_{ij}S_{,ij}\left<\delta\theta_i\delta\theta_j\right>
= S(\bar{\Theta})+\frac{1}{2}\sum_{ij}S_{,ij}(F^{-1})_{ij}\,,
\eeq
where we utilize the relation ${\rm Cov}(\theta_i,\theta_j)\equiv\left<\delta\theta_i\delta\theta_j\right> = F^{-1}$ in the second equality.

Then the 1-$\sigma$ variance of $S(\Theta)$ is 
\bea
\sigma_S^2&\equiv& \left<(S(\Theta)-\left<S(\Theta)\right>)^2\right>\no \\
&=& \left<\left(\sum_iS_{,i}\delta\theta_i+\frac{1}{2}\sum_{ij}S_{,ij}\delta\theta_i\delta\theta_j-\frac{1}{2}\sum_{lm}S_{,lm}(F^{-1})_{lm}\right)^2\right> \no\\
&=&\sum_{ij}S_{,i}S_{,j}\left<\delta\theta_i\delta\theta_j\right>+\frac{1}{4}\left[\sum_{ij}\sum_{lm}S_{,ij}S_{,lm}\left<\delta\theta_i\delta\theta_j\delta\theta_l\delta\theta_m\right>\right.\no\\
&&\left.-2\sum_{ij}\sum_{lm}S_{,ij}S_{,lm}\left<\delta\theta_i\delta\theta_j\right>(F^{-1})_{lm}+\sum_{ij}\sum_{lm}S_{ij}S_{lm}(F^{-1})_{,ij}(F^{-1})_{,lm}\right]\no\\
&=& \sum_{ij} S_{,i}S_{,j}(F^{-1})_{ij} + \frac{1}{2}\sum_{ij}\sum_{lm}S_{,ij}S_{,lm}(F^{-1})_{il}(F^{-1})_{jm}\,.
\label{eq:S_fisher2}
\eea
Here the Wick's theorem $\left<\delta\theta_i\delta\theta_j\delta\theta_l\delta\theta_m\right> = \left<\delta\theta_i\delta\theta_j\right>\left<\delta\theta_l\delta\theta_m\right>+\left<\delta\theta_i\delta\theta_l\right>\left<\delta\theta_j\delta\theta_m\right>+\left<\delta\theta_i\delta\theta_m\right>\left<\delta\theta_j\delta\theta_l\right>$ is applied.

If we truncate at the first order of Taylor expansion, 
\beq
\sigma_S^{(1)}=\sqrt{\sum_{ij} S_{,i}S_{,j}(F^{-1})_{ij}}\,.
\eeq
The comparison of $\sigma^{(1)}_S$ and $\sigma^{(2)}_S$ is shown in figure~\ref{fig:sigma_S_compare}.  In the case of varying $\log_{10} M_{\rm c}$, $\theta_{\rm ej}$ and $\delta$, the convergence is not perfect that we  see some discrepancies between $\sigma^{(1)}_S$ and $\sigma^{(2)}_S$ at small scales, yet it is still safe to say that the constraints on $S(k)$ are below $1\%$ at all considered scales. In the case of varying only $\log_{10} M_{\rm c}$ and $\theta_{\rm ej}$, the convergence is manifest and $\sigma_S(k)$ is below $0.1\%$ at all considered scales.

In conclusion, the truncation at the second order of the Taylor expansion in eq.~(\ref{eq:S_taylor}) is qualitatively accurate enough for discussions and conclusions of this work. We will adopt eq.~(\ref{eq:S_fisher}) to calculate $\sigma_S(k)$ in this work.

\section{Sensitivity of $D_{\rm BCM}$ on BCM parameters} 
\label{app:sensitivity_D_BCM}
\begin{figure*}[t!]
\begin{center}
\includegraphics[width=\columnwidth]{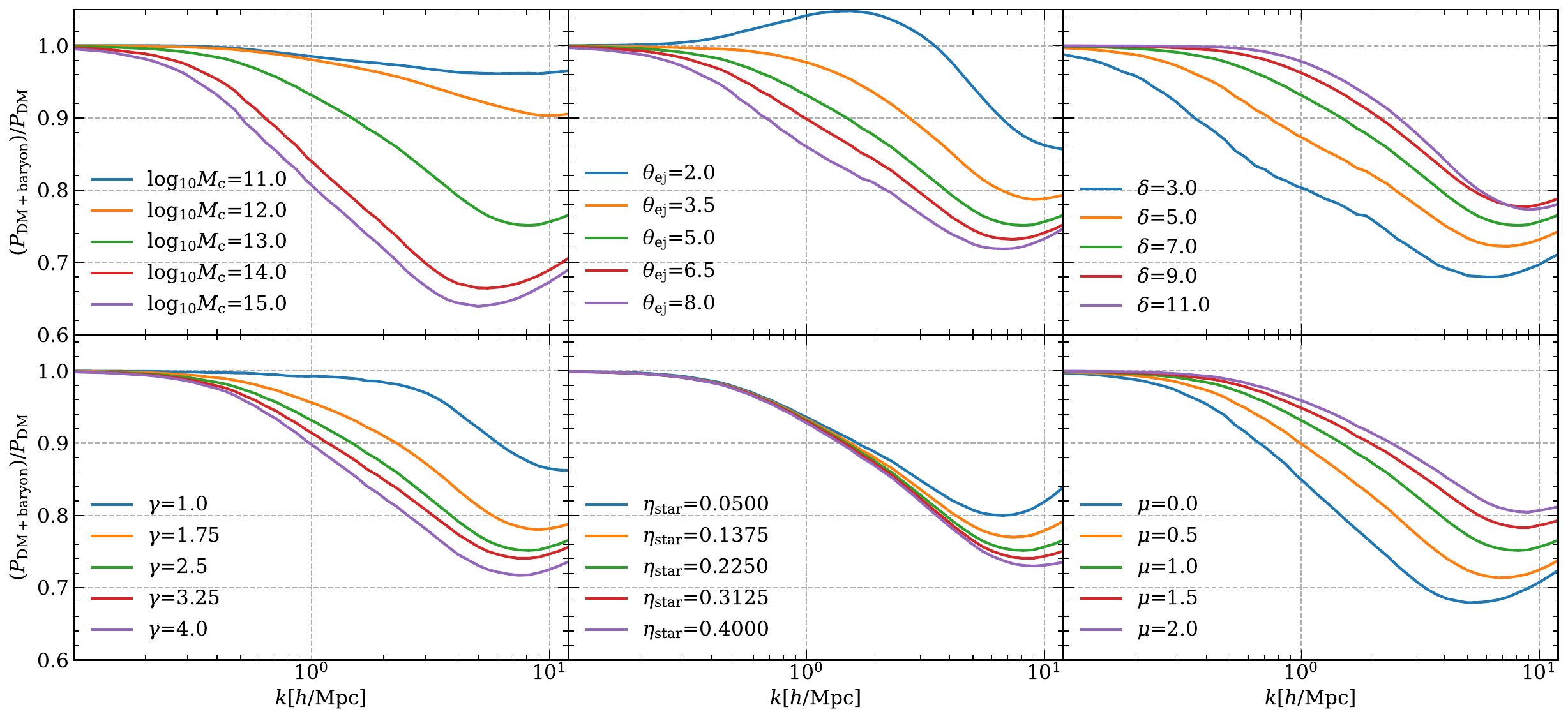}
\end{center}
\caption{The sensitivity test of the matter power spectrum damping with respect to 6 BCM parameters. \ZY{We vary only one parameter at one time and fix other ones to be $\Theta_{\rm mid}$.} We use the BCM emulator developed in~\cite{Giri2021} (\texttt{BCemu} emulator). }
\label{fig:damping_sensitivity}
\end{figure*}
6 BCM parameters in \texttt{S19} model are input parameters of the BCM emulator for $S(k)$ developed in~\cite{Giri2021} (\texttt{BCemu} emulator). We vary only one parameter at one time and fix other ones to be $\Theta_{\rm mid}$. The corresponding  matter power spectrum dampings are shown in figure~\ref{fig:damping_sensitivity}. We see that $S(k)$ is sensitive to all parameters except $\eta_{\rm star}$, and the degeneracy between parameters is obvious.

\section{The $\mu$ dependency of the $S(k)$ constraints}
\label{app:mu_dependency}
\begin{figure*}[t!]
\begin{center}
\includegraphics[width=0.6\columnwidth]{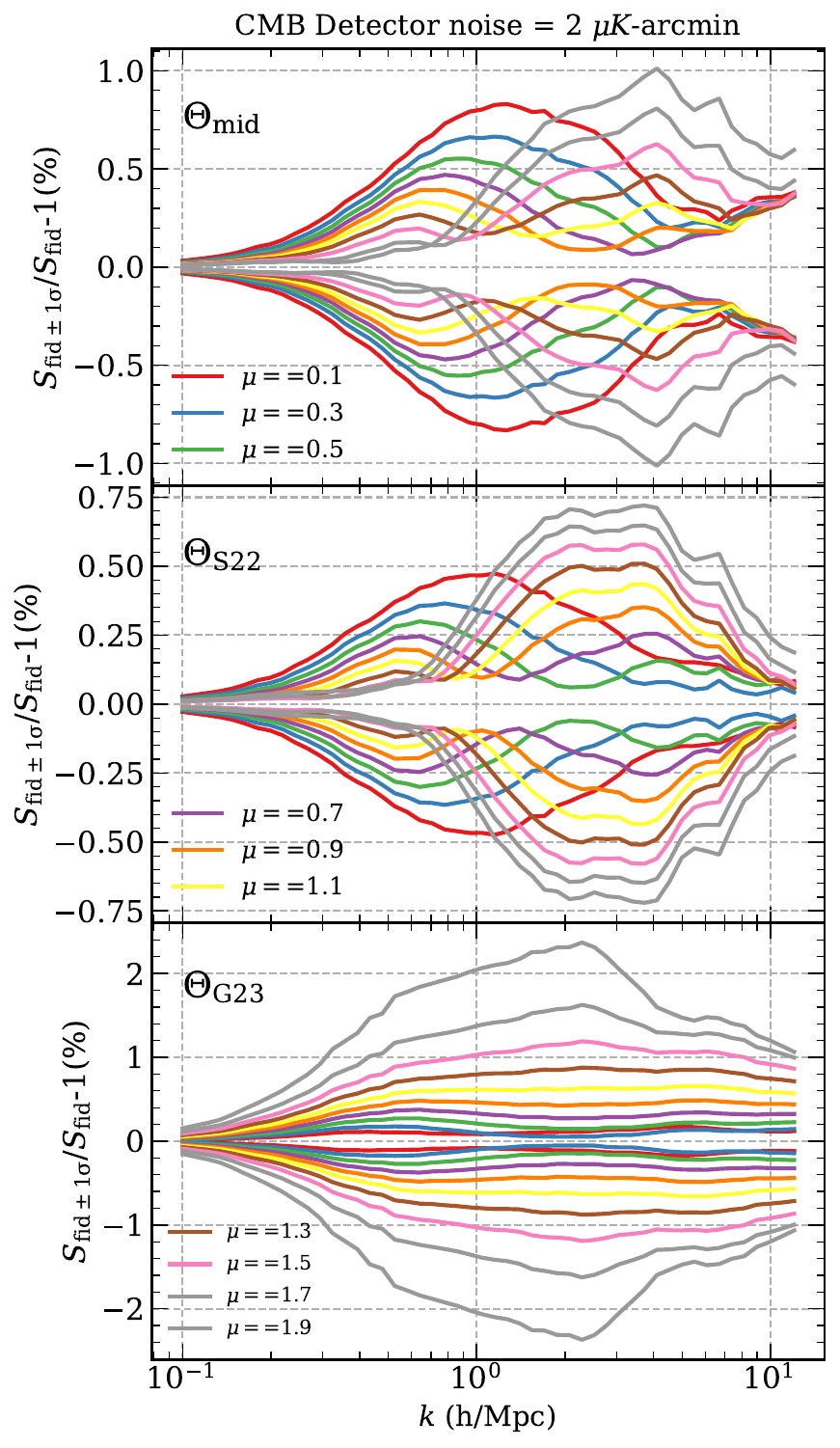}
\end{center}
\caption{\ZY{The $\mu$ dependency of the $S(k)$ constraints when 3 parameters are varied. We only vary $\mu$ and fix other BCM parameters to be those of $\Theta_{\rm mid}$, $\Theta_{\rm S22}$, or $\Theta_{\rm G23}$. Here the CMB experiment is assumed to have a detector noise of $2\mu K\mathchar`-{\rm arcmin}$.}}
\label{fig:mu_dependency}
\end{figure*}
\ZY{Taking the $\mu$ variation as an example, figure~\ref{fig:mu_dependency} shows how the $S(k)$ constraints depend on the fiducial BCM parameter values adopted. Both the amplitude and scale dependency of the constraint depends on the fiducial $\Theta$. A thorough study of this $\Theta$ value dependency needs considering more variations in the parameter space, which is left to the future work.}

\bibliographystyle{JHEP}
\bibliography{mybib}







\end{document}